\documentclass[a4paper,11pt]{article}
\pdfoutput=1 

\usepackage{tikz}
\usetikzlibrary{shapes.misc}
\tikzset{cross/.style={cross out, draw=black, fill=none, minimum size=2*(#1-\pgflinewidth), inner sep=0pt, outer sep=0pt}, cross/.default={2pt}}

\usepackage[T1]{fontenc} 

\usepackage{jheppub}

\def\ket#1{|{#1}\rangle}
\def\bra#1{\langle{#1}|}
\def\Tr{\text{Tr}}

\newcommand{\comment}[1]{}

\title{\boldmath Defects and Perturbation}

\author{Enrico M. Brehm}

\affiliation{
	Max Planck Institut f\"ur Gravitationsphysik,\\
	Albert-Einstein-Institut, \\
	Potsdam-Golm, D-14476, 
	Germany}

\emailAdd{brehm@aei.mpg.de}

\abstract{We investigate perturbatively tractable deformations of topological defects in two-dimensional conformal field theories. We perturbatively compute the change in the $g$-factor, the reflectivity, and the entanglement entropy of the conformal defect at the end of these short RG flows. We also give instances of such flows in the diagonal Virasoro and Super-Virasoro Minimal Models.}

\begin{document} 
\maketitle
\flushbottom


\section{Introduction}
\label{sec:intro}

A \textit{defect} in a two-dimensional conformal field theory is given by a line where the fields of the theory are glued together in a specific way. 
A defect is called \textit{conformal} when the gluing condition is compatible with all conformal symmetry transformations that leave the defect line invariant.
It is clear that most defects are not conformal and, in particular, are not invariant under scale transformations. 
This leads to a renormalization group flow in the space of defects. The fixed points of these flows are generically conformal.

In physics conformal defects appear in various situations, with very different effects and interpretation. 
They appear very naturally in two-dimensional statistical systems and $(1+1)$ dimensional quantum systems. 
Consider for example a two-dimensional lattice model where couplings are altered from their normal values along a line~\cite{Delfino:1994nx}. 
In the continuum limit this gives rise to a defect. 
Another physical situation where defects arise naturally is the junction of quantum wires \cite{Claudio2003} in the continuum limit. Defects can also be used to describe tunneling in the quantum Hall effect \cite{2009AnPhy.324.1547F}. One can also interpret defects as ``symmetries'' that relate the features of the two theories they connect. They can for example reflect group symmetries or dualities of a theory~\cite{Frohlich:2004ef}, there is a preferred defect that connects the endpoints of renormalization group flows encoding the information of the flow~\cite{Brunner:2007ur,Gaiotto:2012np}, and they can reflect a spectrum generating symmetry in string theory~\cite{Bachas:2008jd}. 

The construction and study of general conformal defects is equivalent to the study of boundary conditions of a folded theory~\cite{Wong:1994np,Oshikawa:1996ww,Oshikawa:1996dj}.
Even for rational conformal field theories the folded model is typically not rational with respect to the symmetry that is preserved by the defect.
Therefore, only in very particular models one can classify conformal defects. 
Among these are the Lee-Yang model~\cite{quella_reflection_2007} and the critical Ising models~\cite{Oshikawa:1996dj}, and a discussion of supersymmetry preserving defects in the tricritical Ising model can be found in \cite{Gang:2008sz,Makabe:2017ygy}. 
Large classes of conformal defects can be constructed between free theories~\cite{Bachas:2001vj,Bachas:2012bj}. 
In more general theories there exist two subclasses of defects that are much easier to describe and classify: topological and factorizing ones, that will be discussed in section \ref{sec:TopDefects}.

In the present paper we study some selected aspects of defects which are almost topological by means of perturbation theory. 
The perturbative analysis of an RG flow can provide candidates for non-topological conformal defects and allows the calculation of properties along the flow. 
We consider perturbations by almost marginal fields on topological defects and in particular on the identity, such that the flow remains `short' and the perturbative analysis provides reliable results.

Constructing conformal defects from perturbations along a line can be seen as a specialisation of the boundary perturbation which has been introduced first in~\cite{Ghoshal:1993tm}.
For an early example of a construction of topological defects by a chiral perturbation see {\it e.g.}~\cite{Konik:1997gx}; a more recent one is {\it e.g.}~\cite{Runkel:2007wd}.  Our work is particularly inspired by~\cite{Kormos:2009sk,Makabe:2017uch}, where a two-parameter chiral+antichiral perturbation on a topological defect of the Virasoro Minimal Models leads to a particular instance of a conformal defect which is neither factorizing nor topological. A further recent motivation to search for tractable instances of pure defect flows was~\cite{Karch:2018uft}. We also want to mention the interesting approach in \cite{Budzik:2020aqg} where the authors explore the space of conformal defects by using open string field theory techniques. 

In order to find potential examples for our perturbation, we screen the spectrum of elementary topological defects in Virasoro Minimal Models \cite{Cappelli:2009xj} and in $\mathcal{N}=1,2$ Super-Virasoro Minimal Models~\cite{Friedan:1984rv,Eichenherr:1985cx,Bershadsky:1985dq,Goddard:1986ee,DiVecchia:1985ief} for candidate perturbations of short flows. 
The list we obtain is not exhaustive, since our criterion will be simply that the set of operators obtained from the initial perturbation by successive OPE only contains almost-marginal operators. 
This naive condition excludes situations where more relevant operators are obtained from successive OPE, but nevertheless not required as counterterms. 
The candidates we find then potentially generate a larger spectrum of defects by fusion with topological defects~\cite{Bachas:2008jd}.  

In explicit perturbative calculations we focus on spin-less one-parameter deformations by operators of conformal dimension $\Delta=1-\delta$, with small $\delta$, in the spectrum of the identity defects.
We compute the change in the $g$-factor~\cite{Affleck:1991tk}, the reflection coefficient~\cite{quella_reflection_2007}, 
and the entanglement entropy through the defect~\cite{sakai_entanglement_2008}. 
Our results for the former two quantities hold to third order in the perturbation parameter $\delta$, 
while the entanglement entropy is computed to second order. 
The $g$-factor result obeys the $g$ theorem~\cite{Friedan:2003yc,Casini:2016fgb}, and in particular remains at the same value for exactly marginal deformations. The initial topological defects are totally transmissive, and reflectivity
increases for all relevant perturbations except for those which are chiral ({\it i.e.}, which transform trivially under either 
the holomorphic or antiholomorphic copy of the Virasoro algebra). In case of the entanglement entropy, the expectation is that
correlations across a defect should be reduced by the perturbation. Degrees of freedom in the bulk of the theory are unchanged, and degrees of freedom on the defect should only disturb correlations across the defect. Indeed, to first order in perturbation the leading order term in the entanglement entropy is reduced. Defects of the Ising model suggest a relation between the changes of entanglement entropy and reflectivity~\cite{Brehm:2015lja}. Our computations show that the relation to leading order is the
same as in the Ising models for all spin-less one-parameter flows.

The text is organized as follows. In section \ref{sec:confDef} we briefly recall the aspects of conformal defects relevant 
for our analysis, and introduce the three properties that we analyse perturbatively later. Section~\ref{sec:defectPurt} 
details generalities of the perturbation, and we conduct the search for candidate fields in the Minimal Models. 
In section~\ref{sec:pertfus} we comment on the implications of fusion with topological defects. Finally, in section \ref{sec:pertAnalysis} we perform the perturbative calculations for the $g$-factor, the reflectivity, and entanglement entropy on the identity. 


\section{Conformal defects}\label{sec:confDef}
For a defect along the real axis in the complex plane, the local condition of preserving conformal symmetry transformations is
\begin{equation}\label{eq:gluingcond}
 \lim_{~y\to0^+}\left( T^{(1)}(x+iy)-\tilde T^{(1)}(x-iy)\right) 
 = \lim_{~y\to0^-}\left( T^{(2)}(x+iy)-\tilde T^{(2)}(x-iy)\right) \,,
\end{equation}
where $T^{(i)}$ and $\tilde{T}^{(i)}$ are the holomorphic and antiholomorphic components of the energy momentum tensor of the theory defined on the upper ($i=1$) and lower ($i=2$) half-plane. 
For time evolution orthogonal to the defect line, the defect acts as an operator $\mathcal{I}$ which 
maps states from the Hilbert space of one theory to the other. The above gluing condition then 
turns into the statement that this operator satisfies the relation
\begin{equation} \label{eq:gluing}
 \left(L_m^{(1)}-\bar L_{m}^{(1)}\right)\mathcal{I} = \mathcal{I}\left(L_m^{(2)}-\bar L_{m}^{(2)}\right)
\end{equation}
with the generators of the conformal transformations.
Besides this local condition, defects have to satisfy a Cardy condition, which ensures that quantization parallel and quantization orthogonal to the defect are equivalent. Quantization parallel to the defect gives rise to the spectrum of defect (changing) fields. 

\subsection{Factorizing and topological defects}\label{sec:TopDefects}

The gluing condition \eqref{eq:gluing} has two special solutions where its treatment simplifies considerably. 
The first one occurs when both sides of the equation vanish, which yields separate 
boundary conditions for the two conformal field theories. The corresponding defect is called \textit{factorizing}. 

The other special solutions occur if the holomorphic and the antiholomorphic parts of the energy momentum tensor are separately 
continuous across the defect,
\begin{equation}
L_{m}^{(1)}\mathcal{I} = \mathcal{I} L_m^{(2)}\qquad\text{and} \qquad \bar L_{m}^{(1)}\mathcal{I} = \mathcal{I} \bar L_m^{(2)}\qquad\forall\, m\in\mathbb{Z} \,.
\end{equation}  
In this case the defect is compatible with all conformal transformations. It commutes in particular with the Hamiltonian and the momentum operator and, hence, is tensionless and can be moved around without cost of energy or momentum. Such defects can be deformed arbitrarily on the surface without changing correlation functions, as long as they do not cross any operator insertion. This type of defect is called \textit{topological}~\cite{Petkova:2000ip}. Since $T$ and $\tilde{T}$ are separately continuous across the defect, the spectrum of defect fields is organized in products of representations of the holomorphic and anti-holomorphic copy of the Virasoro algebra. We can therefore attribute a left and a right conformal weight $(h,\bar h)$ to defect fields $\phi$.  

For rational conformal field theories there exists a classification of topological defects~\cite{Petkova:2000ip}. When realized as an operator, a topological defect acts as a constant map between isomorphic representations of the Virasoro or possibly an extended chiral symmetry algebra. 
In case of diagonal theories, {\it i.e.} rational theories which are charge conjugation 
invariant, and where the multiplicities for all chiral algebra representations are~1, the set of elementary
defects preserving the full chiral algebra is given by
\begin{equation}\label{RCFTdefects}
\mathcal{I}_{a}=\sum_{i}\frac{S_{ai}}{S_{0i}}\,\|i\|\,,
\end{equation}
where $\|i\|$ is a Ishibashi-type projector on the representation $\mathcal{H}_i\otimes \overline{\mathcal{H}}_i$, $S_{kl}$ 
are the modular $S$ matrix elements, and 0 labels the identity or vacuum representation. There exists an elementary topological defect for every chiral represention label, or in other words there are as many elementary topological defects in diagonal rational theories as there are primary fields. Due to the Cardy condition, every topological defect can be written as positive integer linear combination of elementary defects. 

The spectrum of defect fields on the fundamental topological defect $\mathcal{I}_{a}$ can also be obtained from the 
Cardy condition. The defect field in the left- and right-moving representations $(j,\bar{\textit{\j}})$ 
appears with multiplicity \cite{Petkova:2000ip}
\begin{equation}\label{eq:defectSpectrum}
M_{j,\bar{\textit{\j}}}=\sum_l N_{al}^a N_{l\bar{\textit{\j}}}^j \,,
\end{equation}
where $N_{ab}^c$ are the fusion rules of the chiral algebra.
Notice that the full bulk spectrum of the diagonal theory is included, since
\begin{equation}\label{eq:bulkINdefectZ}
M_{ii}^{{(a)}} =   \sum_{l} N_{al}^a N_{li}^i \geq N_{a0}^a N_{0i}^i = 1\,.
\end{equation}

\subsection{Defect entropy, reflectivity, and entanglement entropy}

\label{sec:properties}

Now we want to return to general conformal defects and discuss some of their features. 

The entropy $\log g$ of a defect is defined in the same way as the boundary entropy \cite{PhysRevLett.67.161}, and coincides with the boundary entropy in the folded picture. 
In the special case of topological defects \eqref{RCFTdefects} one has
\begin{equation}
g_{\mathcal{I}_{a}} = \frac{S_{a0}}{S_{00}}\,,
\end{equation}
where again $S_{ij}$ are the entries of the modular $S$ matrix. 

Two further quantities defined for all conformal defects are the reflection coefficient $\mathcal{R}$ and the transmission coefficient $\mathcal{T}$ of energy and momentum through the defect \cite{quella_reflection_2007},
\begin{equation}
 \mathcal{R} \equiv \frac{\langle T_1\bar{T}_1 +T_2\tilde{T}_2\rangle_\mathcal{I}}{\langle (T_1+\tilde{T}_2)(\tilde{T}_1+T_2) \rangle_\mathcal{I}}\,,~~~~\mathcal{T} \equiv \frac{\langle T_1\bar{T}_2 +T_2\tilde{T}_1\rangle_\mathcal{I}}{\langle (T_1+\tilde{T}_2)(\tilde{T}_1+T_2) \rangle_\mathcal{I}} \,,\label{eq:ReflTrans}
\end{equation}
where the operator insertions are at the point $z$ in $CFT_1$, and at the corresponding point reflected at the defect in $CFT_2$. For a free boson or fermion, $\mathcal{R}$ and $\mathcal{T}$ are related to the probabilities of reflection and transmission of free field modes. 
The two coefficients satisfy $\mathcal{R}+\mathcal{T}=1$. 
Topological defects are fully transmissive, \textit{i.e.} $\mathcal{T} = 1$ and $\mathcal{R} = 0$, whereas factorizing defects are purely reflective, \textit{i.e.} $\mathcal{R} = 1$ and $\mathcal{T} = 0$. In unitary theories, the respective coefficients of general conformal defects lie between these two extremes. 

Finally, we will be interested in the entanglement entropy through a conformal defect~\cite{sakai_entanglement_2008}. A conformal defect in $(1+1)$ dimensions naturally separates a physical system into two subsystems. Assume a CFT is defined on the complex plane, with a conformal defect spatially fixed at $x_0$ and extending in time direction. Then the system can be spatially separated into $A = (x_0,\infty)$ and $B = (-\infty,x_0)$. The entanglement entropy between these two subregions is what we call the entanglement entropy through the defect. 

In general, if a system is in a pure state $\ket{\psi}$ and its Hilbert space can be decomposed as $\mathcal{H} = \mathcal{H}_A \otimes \mathcal{H}_B$, then the entanglement entropy is defined as the von Neuman entropy of the density matrix reduced to one of the two systems\footnote{see \cite{Calabrese:2009qy} and references therein for a discussion of entanlement entropy in 2d CFT}, 
\begin{align}
E_{A} &= -\Tr_{\mathcal{H}_A} \rho_{A} \log \rho_A = -\Tr_{\mathcal{H}_B} \rho_{B} \log \rho_B =E_B\,,\\
\rho_{A/B} &= \Tr_{\mathcal{H}_{B/A}} \ket{\psi}\bra{\psi}\,.
\end{align} 

\noindent
The entanglement entropy through general conformal defects has been calculated for free bosons in \cite{sakai_entanglement_2008} and for the critical Ising model in~\cite{Brehm:2015lja}. A result for the entanglement entropy through topological defects in rational unitary theories was obtained in~\cite{Brehm:2015plf}.\footnote{We also want to mention that the entanglement entropy through topological defects in holomorphic theories is directly related to the entanglement between left and right moving degrees of freedom at a boundary in the folded theory \cite{Das:2015oha}.} The method in all these cases makes use of the replica trick, which consists of considering $K$ copies of the reduced system and calculating the entanglement entropy from
\begin{equation}
E_{A} = -\lim_{K\to 1} \partial_K \Tr\rho_A^K\,.
\end{equation}

\begin{figure}[t]
 \begin{center}
  \begin{tikzpicture} [scale=2]
   \fill [gray!20] (-.4,1.1) rectangle (0,-1.1);
   \draw[->] (-.4,0) -- (1.1,0) node[below] {\small{Re $w$}};
   \draw[->] (0,-1.1) -- (0,1.1) node[above] {\small{Im $w$}};
   \draw[blue,very thick] (0,0) -- node[above] {\small{branch cut}} (1,0);
   \draw[very thick,red] (0,-1) -- (0,0);
   \draw[very thick,red] (0,0) -- node[right] {\small{Interface}} (0,1);
   \node at (1.7,0) {$\longrightarrow$};
  \end{tikzpicture}
  \begin{tikzpicture} [scale=3]
   \foreach \j in {1,...,5} \fill[gray!20] (-1,2*.\j) rectangle (1,2*.\j-.1);
   \foreach \i in {1,...,9} \draw[red] (-1,.\i)--(1,.\i);
   \draw[->] (-1.1,0) -- (1.1,0) node[right] {\small{Re $z$}};
   \draw[->] (0,-.2) -- (0,1.1) node[above] {\small{Im $z$}};
   \draw[blue] (-1,0) node[below] {\small{$\log \epsilon$}} --  (1,0) node[below] {\small{$\log L$}} -- (1,1) node[right] {\small{$2\pi K$}}-- (-1,1) -- (-1,0);
  \end{tikzpicture}
 \end{center}
 \caption{Sketch of the $K$-sheet Riemann surface $\mathcal{R}_K$ that one needs for the replica trick. After imposing a IR citoff $\epsilon$, an UV cutoff $L$, and suitable variable transformation, $\mathcal{R}_K$ looks like on the right. }
 \label{fig:InterfaceAndSheet}
\end{figure}
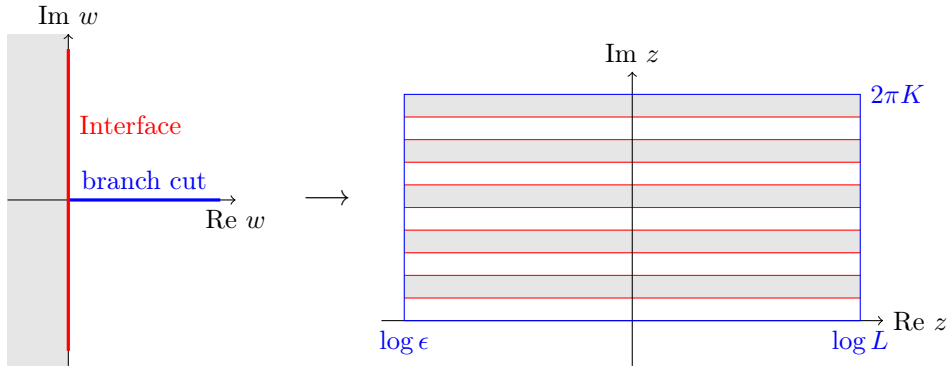

\noindent 
The right-hand side can also be given in terms of a partition function on a $K$-sheeted Riemann surface $\mathcal{R}_K$, $\Tr\rho_A^K = \frac{Z(K)}{Z(1)^K}$. After introducing cutoffs $\epsilon$ and $L$ in the ultraviolet and the infrared respectively, this partition function is equivalent to the partition function on a torus with $2K$ defect insertions,
\begin{equation}
Z(K) = \Tr \left(\mathcal{I}^\dagger e^{-\delta H} \mathcal{I} e^{-\delta H}\right)^K\,,
\end{equation}
where 
$\delta = \frac{2\pi^2}{\log L/\epsilon}$. See figure \ref{fig:InterfaceAndSheet} borrowed from \cite{Brehm:2015lja} for a sketch of the construction. The entanglement entropy is then given in terms of the free energy $F(K) = \log Z(K)$ by
\begin{equation}\label{eq:EEfromF}
E = \lim_{K\to 1} (1-\partial_K) F(K)\,.
\end{equation}


\section{Defects from perturbation}\label{sec:defectPurt}
\label{sec:pert}
For topological defects ${\cal I}$ along the line $\gamma_\mathcal{I}$, we consider perturbations by defect 
fields which transform as Virasoro primary states with weights $h$ and $\bar{h}$. The perturbation is relevant
if the conformal dimension $\Delta = h+\bar h < 1$.

Since the spectrum of all topological defects includes the bulk spectrum, relevant defect perturbations 
generically exist if the bulk spectrum contains non-trivial operators of conformal dimension $\Delta<1$.
This holds in particular for the identity defect.

For a generic perturbation by relevant or marginal defect operators $\phi_i$, the 
action of the system is perturbed by a term 
\begin{equation}\label{eq:genericperturbation}
\delta S =  \sum_i\lambda_i \int_{\gamma_\mathcal{I}} dw \,\phi_i(w)\,,
\end{equation}
where $\lambda_i = \tilde{\lambda_i}\varepsilon^{\Delta_i-1}$ is the coupling. The length scale $\varepsilon$ is part of the 
cutoff scheme and $\tilde{\lambda}$ is the dimensionless coupling constant. 

For perturbation theory to be applicable we will assume that all 
anomalous dimensions $\delta_i=1-\Delta_i\geq 0$ which appear in the 
sum~\eqref{eq:genericperturbation} are small, such that an expansion in 
the $\delta_i$ makes sense. Furthermore, we will assume that the $\phi_i$ form a 
closed set, in the sense that the fusion orbit or ``OPE family'' of the 
$\phi_i$ --- {\it i.e.} the set of operators generated through repeated operator 
product expansion --- contains no non-trivial relevant or marginal fields other than 
the $\phi_i$ in \eqref{eq:genericperturbation}. 

The expectation value of an operator $\mathcal{O}(z,\bar{z})$ 
in the presence of the perturbed defect is given by
\begin{equation}\label{eq:pertCorrelation}
 \langle\mathcal{O}(z) \mathcal{I}_\text{pert.}\rangle =  \langle\mathcal{O}(z) \mathcal{I}\rangle + \langle\mathcal{O}(z) \mathcal{I}\delta S\rangle + \tfrac{1}{2}\langle \mathcal{O}(z) \mathcal{I}(\delta S)^2\rangle + \dots\,.
\end{equation}
The system of beta functions up to third order in the renormalized coupling 
constants~$\lambda_i$ reads
\begin{equation}\label{eq:betaOPEscheme}
\beta_i=\delta_i\lambda_i-\sum_{jk}{\cal C}_{jk}^i\lambda_j\lambda_k+\sum_{jkl}{\cal D}_{jkl}^i\lambda_j\lambda_k\lambda_l\,.
\end{equation}
For non-vanishing coefficients ${\cal C}$ and ${\cal D}$ of order 1, and all $\delta_i\sim\delta_j$ 
of the same (small) order, there can exist a non-trivial IR fixed point of the perturbed system,
where the values $\lambda_i^*$ of the renormalized coupling constants are
of order $\delta_i$. If the perturbation~\eqref{eq:genericperturbation} 
contains only a single operator $\phi$,
\begin{equation}\label{eq:lC}
\lambda^* =  \frac{\delta}{{\cal C}} - \frac{{\cal D} \delta^2}{{\cal C}^3} + \mathcal{O}(\delta^3)\,.
\end{equation}
In cases where the coefficient ${\cal C}$ vanishes, {\it i.e.} where
\begin{equation}\label{eq:betaCvanishes}
 \beta = \delta \lambda + {\cal D} \lambda^3 + \mathcal{O}(\lambda^4)\,,
\end{equation} 
there can generically only be a short flow if ${\cal D}<0$, in which case
$\lambda^*$ would be of order $\sqrt{\delta}$,
\begin{equation}\label{eq:lC0}
\lambda^* = \pm \frac{\sqrt{\delta}}{\sqrt{{-\cal D}}} + \mathcal{O}(\delta)\,.
\end{equation}
In the calculation of the coefficients in the beta functions we employ 
the so-called OPE scheme. In this scheme,
\begin{equation}
{\cal C}_{jk}^i=C_{jk}^i
\end{equation}
for the OPE coefficients
\begin{equation}\label{eq:selfOPE}
 \phi_j(x)\phi_k(y) = \frac{\delta_{jk}}{(x-y)^{2h_j}(\bar{x}-\bar{y})^{2\bar{h}_j}} + 
\sum_i \frac{C_{jk}^i \phi_i(y) }{(x-y)^{-h_i+h_j+h_k}(\bar{x}-\bar{y})^{-\bar{h}_i+\bar{h}_j+\bar{h}_k}}
+\ldots\,,
\end{equation}
where we omitted descendent terms.
In the situation which will be of interest to us in section~\ref{sec:pertAnalysis}, the coefficient ${\cal D}$ can be written in terms of an integral over the cross ratio $\eta = \frac{z_{12} z_{34}}{z_{13}z_{24}}$ by adapting a result of~\cite{Gaberdiel:2008fn},
where an OPE-cutoff minimal subtraction scheme is employed:
\begin{align}\label{eq:D}
\quad {\cal D}_{jkl}^i=\frac{1}{6}
\int_0^1 d\eta\,\sum_{{\rm perm}(j,k,l)}\bigg(&
\eta^{-\frac{2}{3}}(1-\eta)^{-\frac{2}{3}}\,Y_{jk,li}(\eta) \\
&\quad-\sum_p\left(C_{jk}^pC_{pl}^i\,\eta^{\Delta_p-2}+C_{kl}^pC_{jp}^i\,(1-\eta)^{\Delta_p-2}\right)\bigg)\,.\nonumber
\end{align}
In the last expression, $Y$ is part of the four-point correlation function,
\begin{equation}
Y_{jk,li}(\eta)=\sum_pC_{jk}^pC_{lp}^i F_{jk,li}^p(\eta)\tilde{F}_{jk,li}^p(\bar{\eta})
\end{equation}
where $F_{jk,li}^p(\eta)$ is a conformal block in the channel where
$\phi_j$ approaches $\phi_k$, and $\phi_l$ approaches $\phi_i$,
with the normalization
\begin{equation}
F_{jk,li}^p(\eta)=\eta^{h_p-\frac{2}{3}}\,,\quad 
\tilde{F}_{jk,li}^p(\bar{\eta})=\bar{\eta}^{\bar{h}_p-\frac{2}{3}}\qquad{\rm as}\quad \eta\rightarrow 0\,. 
\end{equation}
The sum in the second line of \eqref{eq:D} runs over all Virasoro conformal primary fields
which appear in the spectrum of the original topological defect, and have $\Delta_p\leq 1$.

\subsection{(Anti-)holomorphic and spinless perturbations}\label{sec:HolPert}

Two types of perturbations on a topological defect are naturally special.
The first type occurs if the spectrum on a topological defect contains 
(anti-)holomorphic relevant fields.  The corresponding 
(anti-)holomorphic perturbation is believed 
to flow to another topological defect \cite{Konik:1997gx,Runkel:2007wd}. 
Evidence for this comes from the perturbative analysis of the reflection coefficient 
$\mathcal{R}$ as defined in \eqref{eq:ReflTrans}: For a holomorphic 
perturbation $\phi(z)$,
\begin{align}
\langle T(w)\bar{T}(\bar{w}) \,e^{\lambda \int_{\gamma_I} dz \phi(z)}\rangle_\mathcal{I} = \sum_{n=1}^\infty \frac{\lambda^n}{n!} \int_{\gamma_\mathcal{I}^n}  dz_1\dots dz_n \,\langle T(w) \bar{T}(\bar{w}) \phi(z_1)\dots \phi(z_n)\rangle_\mathcal{I}\,. 
\end{align}
The correlation function factorizes into holomorphic and 
antiholomorphic parts, since the topological defect $\mathcal{I}$ does 
not lead to any mixing. It follows immediately that every summand vanishes 
because of the anti-holomorphic insertion $\bar{T}(\bar{w})$\,.
The reflectivity does therefore not change during a holomorphic flow, 
and the conformal defects at the fixed points will have vanishing 
reflection coefficient. An analogous statement holds for anti-holomorphic 
perturbations.

On the identity, {\it i.e.} when there is no defect in the first place,
the fact that such chiral perturbations flow to topological defects
was shown in~\cite{Konik:1997gx}.

A combination of a holomorphic and an anti-holomorphic perturbation of a topological defect has been investigated in~\cite{Kormos:2009sk,Makabe:2017uch}. 
The IR fixed point of the corresponding flow is a perturbatively tractable conformal but non-topological defect.

The second type of perturbations in which we will be interested 
are perturbations by a defect operator with $h=\bar{h}$. Such a perturbation
generically triggers both chiral and non-chiral fields, but 
there are classes of topological defects which do not contain
any more relevant fields in their spectrum --- the primary
example for this is the identity defect. For each of these 
spinless perturbations, an operator of the same dimension 
also appears in the bulk spectrum of our diagonal theories.  

\subsection{Candidate fields in some simple RCFT coset models}
\label{sec:search}
In the following we analyze the (defect) field content of diagonal minimal models to deduce when to expect the existence of almost topological conformal defects. The three series we will consider are 

\begin{itemize}
\item The Virasoro Minimal Models which can be realized by the coset \cite{DiFrancesco:1997nk}
\begin{equation}\label{eq:CosetN0}
\frac{\mathfrak{su}(2)_k\oplus\mathfrak{su}(2)_1}{\mathfrak{su}(2)_{k+1}}\,, \qquad k \in \mathbb{N}.
\end{equation}
\item The $\mathcal{N}$=1 Super-Virasoro Minimal Models with a coset realization \cite{Zamolodchikov:1988nm,AlvarezGaume:1991bj,Goddard:1984vk}
\begin{equation}\label{eq:CosetN1}
\frac{\mathfrak{su}(2)_k\oplus\mathfrak{su}(2)_2}{\mathfrak{su}(2)_{k+2}}\,, \qquad k\in \mathbb{N}.
\end{equation}
\item The $\mathcal{N}$=2 Super-Virasoro Minimal Models with the coset \cite{DiVecchia:1986fwg,Goddard:1984vk}
\begin{equation}\label{eq:CosetN2}
\frac{\mathfrak{su}(2)_k\oplus\mathfrak{u}(1)_2}{\mathfrak{u}(1)_{k+2}}\,,\qquad k \in \mathbb{N}.
\end{equation}
\end{itemize}

In all theories we will consider the large $k$ regime and the finite $k$ regime seperately. 

\subsubsection*{Virasoro Minimal Models}

We consider the diagonal minimal series realized by the coset \eqref{eq:CosetN0}. 
If we indicate the primary fields of such a model by the Kac labels  $(p,q)$ with $1\le p\le k+1$ and $1\le q \le k+2$, the list of conformal weights of primary fields is given by 
\begin{equation}\label{eq:hpq}
h_{p,q} = \bar h_{p,q} = \frac{((k+3)p-(k+2)q)^2-1}{4(k+2)(k+3)}\,. 
\end{equation} 

\noindent
Representations $(p,q)$ and $(k+2-p,k+3-q)$ are identified,  such that each conformal weight appears precisely once.
The fusion coefficients are products of fusion coefficients $N_{pq}^{(l)r}$ of the $su(2)_l$ WZW models at level $k$ and $k+1$,
\begin{equation}\label{eq:FusionVirMin}
 N_{(p_1,q_1),(p_2,q_2)}^{(p_3,q_3)}=N_{p_1,p_2}^{(k)p_3}\,N_{q_1,q_2}^{(k+1)q_3}\,,
\end{equation}
where $N_{pq}^{(l)r}=1$ if $|p-q|+1\leq r \leq{\rm min}(p+q+1,2l+3-p-q)$ and $p+q+r$ is odd, and $N_{pq}^{(l)r}=0$ otherwise.
A non-zero fusion coefficient $N_{ab}^c$ implies a non-vanishing OPE coefficient $C_{ab}^c$ for primary fields. 

We are interested in defect operators with conformal dimension 
\begin{equation}\label{eq:Vircondition}
\Delta_{p,q,\bar{p},\bar{q}}=h_{p,q}+h_{\bar{p},\bar{q}}\lesssim 1\,,\qquad p+\bar{p}{\;}{\rm and}\;q+\bar{q}\;{\rm even}\,.
\end{equation}

\noindent
The parity restriction is a consequence of the $su(2)$ fusion rules and \eqref{eq:defectSpectrum}.
Remember that we also have the constraint that all fields in the OPE family, {\it i.e.} in the set of defect operators obtained by successive OPEs from the initial perturbation should be less relevant than the initial perturbing operator. Also remember from \eqref{eq:defectSpectrum}  that the defect fields with labels $(j,\bar{\textit{\j}}) = ((p,q),(\bar{p},\bar{q}))$ lives on the elementary defect $a= (m,n)$ if
\begin{equation}\label{eq:defSpecVirMin}
    M_{(p,q),(\bar{p},\bar{q})}^{(m,n)} = \sum_{(u,v)} N_{(m,n),(u,v)}^{(m,n)}N_{(u,v),(\bar{p},\bar{q})}^{(p,q)}> 0
\end{equation}

\noindent
{\bf Large $\mathbf{k}$:\quad } Expanding  \eqref{eq:hpq} we get 
\begin{equation}
 h_{p,q} = \tfrac{(p-q)^2}{4} + \tfrac{p^2-q^2}{4k} + \mathcal{O}\left(\frac{1}{k^2}\right)   
\end{equation}
and see that $h_{p,q}\leq 1$ if $-2\leq p-q\leq 1$.
Operators where both $h_{p,q}$ and $h_{\bar{p},\bar{q}}$ are of the form $h_{r,r}$ are all too relevant for the application of perturbation theory. For $p\neq 1$, the fusion family of $\phi_p$ in the $\mathfrak{su}(2)_k$ WZW model contains at least all possible odd labels. The latter two facts give the requirement that at least one of labels $p,q,\bar{p},\bar{q}$ is equal to 1 if we want no more relevant field in the OPE family. Since $h_{p,1}\leq1$ only if $p\leq 3$,
and $h_{1,q}\leq 1$ only if $q\leq 2$, at 
least one of the summands $h_{p,q}$ and $h_{\bar{p},\bar{q}}$ in
\eqref{eq:Vircondition} must
be from the set
\begin{equation}\label{eq:Virdiscriminantset}
h_{1,1}=0\,,\quad h_{1,2}=\tfrac{1}{4}-\tfrac{3}{4(k+3)}\,,\quad 
h_{2,1}=\tfrac{1}{4}+\tfrac{3}{4(k+2)}\,,\quad h_{1,3}=1-\tfrac{2}{k+3}\,.
\end{equation}

\noindent 
We may pick both $h$ and $\bar{h}$ of the perturbing operator from this list. 
Since $\phi_{(1,2)(1,2)}$ and $\phi_{(2,1)(2,1)}$
are too relevant, while $\phi_{(1,3)(1,3)}$ is irrelevant, this observation 
rules out any perturbatively tractable deformation with $h_{p,q}=h_{\bar{p},\bar{q}}$ 
in the large $k$ limit.

For both  $h$ and $\bar{h}$ from the list \eqref{eq:Virdiscriminantset}, 
we furthermore find the two chiral operators $\phi_{(1,1)(1,3)}$ and $\phi_{(1,3)(1,1)}$,
whose combination gives the two-parameter flow studied 
in~\cite{Konik:1997gx,Runkel:2007wd,Kormos:2009sk,Makabe:2017uch}.

Finally, we may pick only one conformal weight, say the left-moving $h$, 
from \eqref{eq:Virdiscriminantset}, and consider any operator with 
$\bar{h}=h_{\bar{p},\bar{q}}$ in the spectrum, where $(\bar{p},\bar{q})$ does not lie 
on the  boundaries of the Kac table.\footnote{All $(r,s)$ on the boundary of the Kac table are either
in the list~\eqref{eq:Virdiscriminantset}, or have $h>1$.} Notice 
that this necessarily means that the defect is not the identity.
Running through the four choices for $h$, we can  draw the following conclusions:

\begin{itemize}
\item If $h=h_{1,1}$, the sum in \eqref{eq:defectSpectrum} respectively \eqref{eq:defSpecVirMin} must contain at least $l=(\bar{p},\bar{q})$ which in particular means that $N_{(u,v)(\bar{p},\bar{q})}^{(u,v)} \neq 0$. Since $(\bar{p},\bar{q})$ is not on the boundary of the Kac table, this means that $N_{(u,v)(3,3)}^{(u,v)} \neq 0$, too. Therefore, because $(\bar{p},\bar{q})$ generically has $(3,3)$ in its OPE orbit, 
the OPE family of the initial perturbation contains the field $\phi_{(1,1)(3,3)}$. This field is too relevant and we can conclude that non of the fields with labels $((1,1),(\bar{p},\bar{q}))$ suits our requirements. 

\item For $h=h_{1,2}$ or $h=h_{2,1}$, any choice of $(\bar{p},\bar{q})$ makes the field either too relevant or irrelevant at large $k$. In addition, its OPE family will generically contain the relevant field with labels $((1,2),(1,2))$ or $((2,1),(2,1))$ respectively. 

\item In the case $h=h_{1,3}$, any allowed choice of $(\bar{p},\bar{q})$ contains in its OPE family  the field $\phi_{(1,3)(3,3)}$ which exists on all topological defects at large $k$ except the identity. It has the dimension
\begin{equation}\label{anotherN0perturbation}
\Delta_{1,3,3,3}=1+\tfrac{2}{k+2}-\tfrac{4}{k+3}<1,
\end{equation} 
and is indeed marginally relevant. Its OPE family
in principle contains the more relevant operators $\phi_{(1,1)(3,3)}$,
$\phi_{(1,3),(1,1)}$, and $\phi_{(1,1),(1,3)}$. However, these more relevant operators do
not exist in the spectrum of the defects ${\cal I}_{(n,1)}$, for $1\leq n\leq k$. These defects
will furthermore not contain any of the operators $\phi_{(1,3)(r,r)}$ with  $r\neq 3$ in their 
spectrum, such that in fact no relevant fields other than $\phi_{(1,3)(3,3)}$ and $\phi_{(3,3)(1,3)}$ 
will appear in the perturbative expansion.
\end{itemize}

\noindent
Hence, our conclusion is that another perturbatively tractable, non-chiral almost marginal perturbation is realised for $\phi_{(1,3)(3,3)}$ and/or $\phi_{(3,3)(1,3)}$ on the topological defects $a=(n,1)$, $n>1$.

\noindent
{\bf Finite $\mathbf{k}$:\quad} At finite values of $k$, deformation by 
$\phi_{(1,2)(1,2)}$ and $\phi_{(2,1)(2,1)}$ may still yield 
qualitatively acceptable numerical predictions for
the quantities discussed in section~\ref{sec:properties} from perturbation theory. 
Both will not trigger more relevant operators.
The naive expectation is that perturbation theory works best for 
$\phi_{(2,1)(2,1)}$, since it has the larger dimension, but this depends heavily 
on the OPE and beta function coefficients.
In the Ising model $k=1$, the representation $(2,1)$ 
is identified with $(1,3)$ at highest weight $h_{1,3}=\tfrac{1}{2}$, 
and the corresponding exactly marginal deformation gives rise to 
three classes of conformal defects continuously connected by RG 
flow to the three fundamental topological 
defects \cite{Oshikawa:1996dj}.

As in the case of large $k$, avoiding strongly relevant perturbations still demands that one of $p,q$ or one of $\bar{p},\bar{q}$ must be 1. Apart from the fields already mentioned, the only other possibility  with $\Delta = 1-\delta$ such that $\delta$ is reasonably small are the fields $\phi_{(1,3)(1,2)}$ and $\phi_{(1,2)(1,3)}$ at levels $k=2,4,6,8$ on the defects $(\frac{k}{2}+1,\frac{k}{2})$ and $(\frac{k}{2}+1,\frac{k}{2}+1)$.\footnote{These labels do not have $s+\bar{s}$ even due to field identifications for the particular values of $k$.}

\begin{boldmath}
\subsubsection*{$\mathcal{N} \!=\! (1,1)$ Super-Virasoro Minimal Models}
\end{boldmath}

\noindent
We consider the diagonal series of supersymmetric models \eqref{eq:CosetN1} for $k\ge 1$. The allowed values for the conformal weights of the superprimary fields are given by
\begin{equation}\label{eq:hN=1}
h_{r,s} = \frac{((k+4)r-(k+2)s)^2-4}{8(k+2)(k+4)} + \frac1{32} (1-(-1)^{r-s})\,,
\end{equation}
for $1\le r < k+1$ and $1\le s < k+3$. Even values of $r-s$ correspond the Neveu-Schwarz sector, and $r-s$ odd to the Ramond sector.
We can identify the labels $(r,s)$ with coset labels $(l,m,n)$ as follows.
For labels $(l,m,n)$ in the standard range
\begin{equation}
0\leq l\leq k,,\qquad 0\leq m\leq 2\,,\qquad 0\leq n\leq k+2\,,
\end{equation}
the selection rule that $l+m+n$ is even and the field identification $(l,m,s)=(k-l,2-m,k+2-n)$ apply. For labels $r,s,l,m,n\ll k$, the Ramond 
representation $(r,s)$ corresponds to the coset representation labels $(r-1,1,s-1)$. 
The Neveu-Schwarz representation with labels $(r,s)$ has two associated coset representations: The representation $(r-1,r-s\,{\rm mod}\,4,s-1)$ contains the super-primary ground state of the multiplet, while the highest weight state of $(r-1,r-s+2\,{\rm mod}\,4,s-1)$ is associated to its superpartner (the $G_{-1/2}$ descendent of the super-primary, if $G$ is the supercurrent).
From the fusion rules of the $su(2)$ WZW model we interpret states in the representations with the label $m=0$ as bosonic, whereas states with the label $m=2$ are fermionic. 
The fusion of Virasoro submodules can be computed from the respective $\mathfrak{su}(2)_l$. For the superconformal  modules given by the labels $(r,s)$, the fusion rules are the same as~\eqref{eq:FusionVirMin}.

Besides the inclusion of the superpartners, the identification of candidates for perturbation theory on the supersymmetry-preserving fundamental topological defects proceeds largely in the same way as for the Virasoro Minimal Models.
We consider an initial perturbation by an operator of dimension $h_{r,s}+h_{\bar{r},\bar{s}}$, or possibly one of its superpartners, with $r+\bar{r}$ and $s+\bar{s}$ even. 
The OPE family of an initial perturbation contains full superconformal modules if the initial operator has $m\neq 0$, but only bosonic operators $m=0$ if itself corresponds to the highest weight state of an $m=0$ coset module.  

\noindent
{\bf Large $\mathbf{k}$:\quad} The exclusion of the very relevant operators of the $r=s$, $\bar{r}=\bar{s}$ modules requires at least one of the labels $r,s,\bar{r},\bar{s}$ to be equal to 1.
This means that either $h$ or $\bar{h}$ (or both) must be from the set
\begin{align}\label{N1susyset}
&h_{1,1}=0\,,\quad h_{1,2}=\tfrac{3}{16}-\tfrac{3}{4(k+4)}\,,
\quad h_{2,1}=\tfrac{3}{16}+\tfrac{3}{4(k+2)} \,,\nonumber\\
& h_{1,3}=\tfrac{1}{2}-\tfrac{2}{k+4}\,,\quad  h_{3,1}=\tfrac{1}{2}+\tfrac{2}{k+2}\,.
\end{align}
If we pick both  $h$ and $\bar{h}$ from this set, we obtain as candidates two chiral almost-marginal deformations which can preserve an ${\cal N}=1$ supersymmetry,\footnote{For the setup of manifestly supersymmetry preserving perturbation theory, see~\cite{Gaberdiel:2009hk}}
\begin{equation}
G_{-\frac{1}{2}}\phi_{(1,3)(1,1)}\,,\qquad
\tilde{G}_{-\frac{1}{2}}\phi_{(1,1)(1,3)}\,,
\end{equation}
and one spinless almost-marginal deformation which would break supersymmetry:
\begin{equation}
\phi_{(1,3)(1,3)}\,.
\end{equation}

\noindent 
From their fusion rules, the former operators behave similarly to the chiral $\phi_{(1,3)(1,1)}$ and $\phi_{(1,1)(1,3)}$ operators in the Virasoro Minimal Models, and might therefore present a supersymmetric version of the conformal defect studied in~\cite{Konik:1997gx,Runkel:2007wd,Kormos:2009sk,Makabe:2017uch}.
The fusion class of the latter operator contains the strongly relevant chiral operators $\phi_{(1,3)(1,1)}$ and $\phi_{(1,1)(1,3)}$ on all defects except the identity. However, from the Virasoro fusion rules it has vanishing $C$ and, hence, only for negative $\mathcal{D}$ \eqref{eq:D} we obtain a flow that does not go beyond the validity of perturbation theory. Unfortunately, we were not able to answer the question whether $\mathcal{D}$ is positive or negative. 

If we only pick the left-moving conformal dimension $h$ from~\eqref{N1susyset}, similar reasoning as around~\eqref{anotherN0perturbation} yields that

\begin{itemize}
\item for $h=h_{1,1}$, the OPE family of the initial perturbation will contain $\phi_{(1,1)(3,3)}$ (notice that the super-primar state of the $(3,3)$ module is bosonic),

\item for $h=h_{1,2}$ ($h_{2,1}$) both the super-primary and its superpartner will eventually trigger $\phi_{(1,2)(1,2)}$ ($\phi_{(2,1)(2,1)}$),

\item for $h=h_{1,3}$, an operator from the module of $\phi_{(1,3)(3,3)}$ will be triggered. The super-primary is too relevant, but its superpartner on the holomorphic side 
\begin{equation}\label{superN1deformation}
G_{-\frac{1}{2}}\phi_{(1,3)(3,3)}\,,\qquad \Delta=1+\frac{2}{k+2}-\frac{4}{k+4}<1
\end{equation}
is bosonic, and therefore will not trigger the primary field.
Potentially one also triggers the more relevant operators $\phi_{(1,1)(3,3)},\phi_{(3,3)(1,1)}$ and $\phi_{(1,3)(1,1)},\phi_{(1,1)(1,3)}$. 
However, they do not exist if we specify the elementary topological defect to have label $(n,1)$, for $n>1$. Since on these defects, analogous restrictions to those stated after~\eqref{anotherN0perturbation} apply, we conclude that~\eqref{superN1deformation}  is a valid candidate for a short flow perturbation.
\end{itemize}

\noindent
{\bf Finite $\mathbf{k}:$\quad} The scan through the possible field in the ${\cal N}=(1,1)$ models at rather low $k$ showed that two more candidates. 

The field $(2,1)$ with $h_{2,1} = \frac{3(6+k)}{16(2+k)}$ has conformal weight pretty close to $\frac{1}{2}$, and the perturbation $\phi_{(2,1)(2,1)}$ will not trigger more relevant fields. 

The other candidate is a two parameter deformation with the fields $((3,1),(1,1))$ and $((1,1),(3,1))$ with conformal dimension $\frac{6+k}{4+2k}$ for $k>1$ on the defects with labels $(a>2,b)$. 

\begin{boldmath}
\subsubsection*{$\mathcal{N} \!=\! (2,2)$ supersymmetric minimal models}
\end{boldmath}

\noindent For the diagonal models based on the coset \eqref{eq:CosetN2}, the bosonic submodules of full superconformal modules at level $k \ge1$ are labelled by triples of integers $(l,m,s)$ with
\begin{align}\label{N2labels}
&0\le l \le k\,,\quad-k-1\le m\le k+2\,,\quad-1\le s\le 2\,,\quad l+m+s = 0 \text{ mod }2\,,
\end{align}
with field identifications $(l,m,s)=(k-l,m+k+2,s+2)$. Full superconformal representations in the Ramond sector are isomorphic to the direct sums $(l,m)\equiv (l,m,-1)\oplus(l,m,1)$ ($l+m$ odd), and full Neveu-Schwarz representations are given by $(l,m)\equiv(l,m,0)\oplus(l,m,2)$ ($l+m$ even).
Notice that ``diagonal'' refers to the labels $(l,m)$ of the superconformal algebra.

For a bosonic subalgebra representation with labels $(l,m,s)$ in the ``standard range'' $|m-s|\le l$, the highest weight is
\begin{equation}\label{N2h}
h_{l,m,s} = \frac{l(l+2)-m^2}{4(k+2)}+\frac{s^2}{8} \,,
\end{equation}
and the $U(1)$ R-current charge of the highest weight state is
\begin{equation}\label{N2q}
q=-\frac{m}{k+2}+\frac{s}{2}\,.
\end{equation}

\noindent
If the labels are not in the standard range, \eqref{N2h} holds up to integers, and~\eqref{N2q} up to even integers.
In the NS sector, the bosonic subrepresentations containing the superconformal primary fields are precisely labeled by $(l,m,0)$ in the standard range.
In particular, chiral primary fields \mbox{($h=q/2$)} have labels $(l,-l,0)$, and anti-chiral primary fields \mbox{($h=-q/2$)} have labels $(l,l,0)$. 

The defects we want to perturb are the elementary topological Cardy defects of the diagonal coset model~\eqref{eq:CosetN2}. 
From the point of view of the bosonic subalgebra, one can write down an elementary defect ${\cal I}_{(L,M,S)}$ of the form~\eqref{RCFTdefects} for every allowed label $(L,M,S)$. 
Similarly to the Minimal Model cases before, for large $k$ the spectrum on such a defect (for small labels) consists of products of representations with $l-\bar{l}$ even, with a maximal distance $l-\bar{l}=2L$. 
The $U(1)$ fusion rules however enforce $m=\bar{m}$ and $s=\bar{s}$. 

Elementary defects of the ${\cal N}=(2,2)$ supersymmetric theory are given by the direct sums
\begin{equation}\label{susydefect}
{\cal I}_{(L,M)} = \tfrac{1}{\sqrt{2}}\left({\cal I}_{(L,M,S)}+{\cal I}_{(L,M,S+2)}\right)\,.
\end{equation}
The defect spectrum of ${\cal I}_{(L,M)}$ consists of tensor products of superconformal representations appearing with integer multiplicities, as can be seen from combining~\eqref{eq:defectSpectrum} with the multiplicities of the mixed descendents, {\it i.e.} the defect changing fields appearing in the trace over ${\cal I}_{(L,M,S+2)}^\dagger {\cal I}_{(L,M,S)}^{\phantom{\dagger}}$.
The multiplicity of the superconformal defect primary $\phi_{(l,m)(\bar{l},\bar{m})}$ on the defect ${\cal I}_{(L,M)}$ is
\begin{equation}
M_{(l,m)(\bar{l},\bar{m})}=\delta_{m\bar{m}}^{(2k+4)}\sum_{l'}N_{Ll'}^{(k)\,L}N_{l'\bar{l}}^{(k)\,l}\,,
\end{equation}
where the Kronecker Delta is understood modulo $2k+4$, and the fusion rules are those of $su(2)_k$.
In particular, the square root prefactor in~\eqref{susydefect} ensures that the NS vacuum in the spectrum of ${\cal I}_{(L,M)}$ appears with multiplicity 1.

\noindent
{\bf Large $\mathbf{k}$:\quad} The large $k$ limit of the ${\cal N}=(2,2)$ models was analyzed in~\cite{Fredenhagen:2012rb}.
We stick to the coset description and treat the label $s$ as a threshold for the conformal weight~\eqref{N2h}. 
Above this threshold, for fixed finite value of $l$, values of $m$ within the standard range yield contributions of order $1/k$. When the values of $m$ become too large to be in the standard range, the necessary field identification yields a weight which rather quickly becomes greater than 1.
The weights then asymptotically approach $|m|/2$ as $|m|$ grows large. 

For large levels $k$, weights of order 1 are therefore obtained within the standard range either due to the threshold provided by $s\neq 0$, or by $l\gg m$. 
In addition there are weights of order 1 when, for fixed $l$, the label $m$ has just passed beyond the standard range. 

Non-trivial correlation functions require overall charge neutrality, and a perturbation with $q\neq 0$ will therefore generically be perturbatively tractable only if a field with  charge $-q$ is also switched on.
The way to achieve this is to consider pairs of conjugate primary fields $\phi_{(l,m,s)(\bar{l},m,\bar{s})},\,\phi_{(l,-m,-s)(\bar{l},-m,-\bar{s})}$ as initial perturbations.
However, if both $l$ and $\bar{l}$ are non-zero (and not equal to $k$), the OPE of these two fields will contain the primary state $\phi_{(2,0,0)(2,0,0)}$ on all defects, which is strongly relevant at large $k$. 
Likewise, for perturbing fields with only $l$ or $\bar{l}$ different from 0, the respective defect spectrum necessarily contains the even more relevant primary fields $\phi_{(2,0,0)(0,0,0)}$ or $\phi_{(0,0,0)(2,0,0)}$ respectively, which will again be obtained in the OPE.

At large $k$, we are therefore restricted to perturbations with labels $l=\bar{l}=0$.
All non-trivial operators we can construct under this condition are irrelevant, except for the R sector fields with $m=\pm1$, and the NS sector fields with $m=\pm 2$. 
The R sector fields have $\Delta=1/4-1/(2k+4)$, such that they are too relevant for perturbation theory to yield good quantitative results. 
However, with conformal dimension $\Delta=1-2/(k+2)$, the NS sector fields $\phi_{(0,\pm2,2)(0,\pm2,2)}$ are within our scope. 
By the field identification $\phi_{(0,\pm2,2)(0,\pm2,2)}=\phi_{(k,\mp k,0)(k,\mp k,0)}$ we observe that these are superconformal (anti-)chiral primaries. 
They exist on all elementary topological defects, and succesive OPEs will only produce additional irrelevant fields.
However, these fields do not contain themselves in their OPE. 
Moreover, as we will show momentarily, at least on the identity defect the coefficients ${\cal D}$ are positive, such that the flow will again take the coupling constants beyond the perturbatively tractable range.

So let us check the two-parameter perturbation by the chiral field $\phi_{-}\equiv \phi_{(k,-k,0)(k,-k,0)}$ and the anti-chiral field $\phi_{+}\equiv \phi_{(k,k,0)(k,k,0)}$ at large $k$. 
Their renormalized coupling constants are denoted $\lambda_{-}$ and $\lambda_{+}$, respectively. 
From the fusion rules 
\begin{equation}\label{fusionchiralfields}
(k,\pm k,0)\times(k,\pm k,0)\sim(k,\pm(k-2),2)\,,\qquad (k,k,0)\times(k,-k,0)\sim (0,0,0)
\end{equation} 
we see that all three-point functions involving these two chiral primary fields vanish, and the only non-vanishing four-point function is \cite{MUSSARDO1989191}
\begin{equation}\label{Mussardo-fourpoint}
\left\langle \phi_{(k,k,0)}(z_1)\phi_{(k,k,0)}(z_2)\phi_{(k,-k,0)}(z_3)\phi_{(k,-k,0)}(z_4)\right\rangle = \frac{1}{|z_{13}z_{24}|^{2-\delta}}
\left|\frac{\eta}{1-\eta}\right|^{2-\delta}\,,
\end{equation}
where $\eta = \frac{z_{12}z_{34}}{z_{13}z_{24}}$ is the conformal cross ratio.
The beta functions are  the coupled equations
\begin{align}\label{betafunctionsExample}
\beta_{-}&=\delta \lambda_-+{\cal D}_{++-}^{-}\lambda_+^2\lambda_-\,,\nonumber\\
\beta_{+}&=\delta \lambda_++{\cal D}_{--+}^{+}\lambda_-^2\lambda_+\,.
\end{align}
The formula for the coefficients~\eqref{eq:D} follows the conventions of~\cite{LEWELLEN1992654,Gaberdiel:2008fn}, for which we find from~\eqref{Mussardo-fourpoint} that
\begin{equation}
Y_{--,++}(\eta)=|\eta^2/(1-\eta)|^{4/3}\,,
\end{equation}
corresponding to the absolute square of the conformal block in the channel $(k,k-2,2)$, or respectively at the square of the vacuum block at $1-\eta$. 
The permutations $Y_{-+,-+}$ and $Y_{+-,-+}$ are obtained by replacing $\eta\rightarrow 1/\eta$ and $\eta\rightarrow 1-\eta$ in the expression for $Y_{--,++}$, respectively.
When we calculate the subtractions, we must bear in mind that the sum over $p$ in \eqref{eq:D} runs over relevant {\it Virasoro} primaries. 
The representation $(k,\pm(k-2),2)$ has $h\rightarrow 2$, and therefore yields an irrelevant operator, while the vacuum representation $(0,0,0)$ contains the Virasoro highest weight vector $J_{-1}\ket{0,0,0}$ of weight $h=1$, which yields two marginal Virasoro operators, besides the vacuum.

Since \eqref{eq:D} is supposed to converge without using any cut-off, we can infer the products of OPE coefficients in the $k\rightarrow\infty$ limit, and find
\begin{equation}
{\cal D}^+_{--+}={\cal D}^{-}_{++-}=\frac{2}{3}\,.
\end{equation}

\noindent
With this result, the system \eqref{betafunctionsExample} has indeed no fixed point within perturbative distance from the identity defect.

\noindent
{\bf Finite $\mathbf{k}$:\quad} The fields $\phi_{(2,0,0)(2,0,0)}$ has conformal dimension $\Delta=4/(k+2)$ which is rather  close to the value $1/2$ for small $k$, without successive OPEs triggering more relevant operators.
In particular, in the Minimal Model with $k=2$ this field is an exactly marginal deformation on the topological defects.  
In fact the $k=2$ theory can be realized as a product of the Ising model (free fermion) with a free (compactified) boson, and this operator corresponds to the $\epsilon$ field of Kac labels $(1,3)$ in the Ising model part.
As in the Ising case \cite{Oshikawa:1996dj}, we expect that the moduli space of defects obtained from this perturbation can be parametrized by transmissivity.  

\subsubsection*{Summary}
\begin{table}[ht]
\centering
\begin{tabular}{|c|c|c|}
\hline
 Min. Model & \multicolumn{2}{|c|}{Perturbing fields} \\
 & low $k$ & large $k$  \\ 
\hline 
Virasoro & \parbox{5.5cm}{\vspace{.3cm}$\bullet$ $\phi_{(2,1)(2,1)}$ on any defect,\\
                $\bullet$ $\phi_{(1,2),(1,3)}$ or $\phi_{(1,3),(1,2)}$ on $\mathcal{I}_{(\frac{k}{2}+1,\frac{k}{2})}$ or $\mathcal{I}_{(\frac{k}{2}+1,\frac{k}{2}+1)}$at $k\le8$ even\vspace{.3cm}}
                & \parbox{6cm}{$\bullet$~$\phi_{(1,3),(1,1)}$ and $\phi_{(1,1),(1,3)}$ on $\mathcal{I}_{(a,b\ge 2)}$\\
                $\bullet$ $\phi_{(1,3),(3,3)}$ or  $\phi_{(3,3),(1,3)}$ on $\mathcal{I}_{(a\ge 2,1)}$
}  \\ 
\hline 
$\mathcal{N}=(1,1)$ & \parbox{5.5cm}{
$\bullet$ $\phi_{(2,1)(2,1)}$ on any defect\\ 
$\bullet$ $\phi_{(3,1),(1,1)}$ and $\phi_{(1,1),(3,1)}$ on $\mathcal{I}_{(a\ge 2,b)}$
} & \parbox{6cm}{\vspace{.3cm}
$\bullet$ $\phi_{(1,3)(1,3)}$ on ${\cal I}_{(1,1)}$ (potenially) \\ 
$\bullet$ $G_{-\frac{1}{2}}\phi_{(1,3)(1,1)}$ and $\tilde{G}_{-\frac{1}{2}} \phi_{(1,1)(1,3)}$ on $\mathcal{I}_{(a,b\ge 2)}$\\ 
$\bullet$ $G_{-\frac{1}{2}}\phi_{(1,3)(3,3)}$ or $\tilde{G}_{-\frac{1}{2}} \phi_{(3,3)(1,3)}$ on $\mathcal{I}_{(a\geq 2,1)}$
\vspace{.3cm}}
\\ 
\hline 
$\mathcal{N}=(2,2)$ & $\phi_{(2,0,0)(2,0,0)}$ & --- 
\\ 
\hline 
\end{tabular} 
\caption{Defect fields in Minimal Models that lead to short flows on the respective topological defects.}
\label{tab:fields}
\end{table}
\noindent Table \ref{tab:fields} contains our conclusions about perturbatively tractable RG  flows on the elementary topological defects. In particular in the $\mathcal{N}=(2,2)$ there seem almost no candidates for deformations that lead to short flows, s.t. a perturbative analysis leads to results that can be trusted at the new fixpoint, i.e. in an almost topological conformal defect. 

\newpage


\subsection{Perturbation and fusion}\label{sec:pertfus}
For topological defects, the fusion process yields an algebra over positive integers~\cite{Petkova:2000ip}. 
The set of elementary topological defects provides a basis of this algebra. 
In case of a diagonal rational theory, where the elementary topological defects are labeled by the primary fields of the CFT, the defect fusion product is exactly as the fusion of conformal families, \textit{i.e.}
\begin{equation}
\mathcal{I}_a \star \mathcal{I}_b = \sum_c {N_{ab}}^c \mathcal{I}_c\,,
\end{equation}
where ${N_{ab}}^c$ are the familiar fusion coefficients. 

This concept of fusion can be generalized in a straightforward way to the case where only one of the two conformal defects is topological. A particular instance of this is the fusion of an elementary topological defect ${\cal I}_a$ with a Cardy boundary state $\ket{b}$, which in a diagonal rational theory yields
\begin{equation}
\mathcal{I}_a \ket{b} = \sum_c {N_{ab}}^c \ket{c}\,.
\end{equation}

\noindent 
As shown in \cite{quella_reflection_2007}, the fusion of a topological defect onto an arbitrary conformal defect does not change its reflectivity. 
However, the process generically changes the original conformal defect. The process of fusion therefore generates new conformal defects with the same reflectivity. 
This is also the case for defects obtained from perturbation. In our analysis of section \ref{sec:search} we obtained perturbatively tractable perturbations which we expect will yield almost-topological defects of the Minimal Models. 
By fusion with topological defects we expect to generate a spectrum of new defects which share the same reflectivity. 
This gives a much larger class of almost-topological defects.

The question remains if fusion with another topological defect and perturbation will in general commute. 
Normally, the spectrum of defect fields as well as correlation functions will be rather different on the fusion product. 
It is not even clear if the initial perturbation does always exist on the fusion product. 
However, at least in case of diagonal rational theories this can be shown, \textit{i.e.} if a field in the representation $(j,\bar{\textit{\j}})$ exists on the fundamental defect $\mathcal{I}_a$ then it exist on the fusion $\mathcal{I}_a \star \mathcal{I}_b$, too. 

To show this, assume the opposite. If the field $(j,\bar{\textit{\j}})$ would not exist on the fusion this would mean that its multiplicity in the respective spectrum of defect fields vanishes, 
\begin{equation}
    M_{(j,\bar{\textit{\j}})}^{a\star b} = \sum_{c,c',l} N_{ab}^c N_{ab}^{c'} N_{cl}^{c'} N_{l\bar{\textit{\j}}}^j = 0\,.
\end{equation}

\noindent
Using Verlindes formula one can show that this is equivalent to
\begin{equation}
    M_{(j,\bar{\textit{\j}})}^{a\star b} = \sum_{m,n,l}  N_{bb}^{n} N_{am}^{n} N_{al}^m N_{l\bar{\textit{\j}}}^j = 0\,.
\end{equation}

\noindent
Every single term in the latter expression is a non-negative integer, \textit{s.t.} it can only vanish if each single term vanishes. We in particular can set $m=a$ and $n=0$ and obtain
\begin{equation}
    0 = \left( N_{bb}^{0} N_{aa}^{0}\right) M^a_{(j,\bar{\textit{\j}})} = M^a_{(j,\bar{\textit{\j}})}\,,
\end{equation}
which shows that the field $(j,\bar{\textit{\j}})$ cannot be part of the spectrum of defect fields on $\mathcal{I}_a$ and, hence, proves the claim. 

However, the existence of the initial perturbation is not enough to show that results from perturbation theory are the same in the original defect and on the fusion product. This is because generically the spectrum of the fusion is richer than the spectrum of the original defect. Defect (changing) fields that are not allowed to appear in the (successive) OPE of the initial field might appear on the fusion. This would in particular change the perturbative results to higher order.

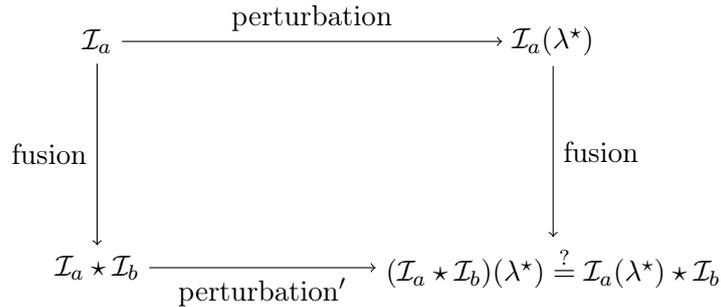
\begin{figure}[t]
\centering
\begin{tikzpicture}[node distance=3cm]
\node (A) {${\cal I}_a$};
\node [right of = A,node distance=6cm] (B) {$\mathcal{I}_a(\lambda^\star)$};
\node [below of = A] (C) {$\mathcal{I}_a\star \mathcal{I}_b$};
\node [right of = C,node distance=6cm] (D) {$(\mathcal{I}_a\star \mathcal{I}_b)(\lambda^\star) \stackrel{?}{=} \mathcal{I}_a(\lambda^\star)\star \mathcal{I}_b  $};
\draw[->] (A) -- node[above]{perturbation} (B);
\draw[->] (A) -- node[left]{fusion} (C);
\draw[->] (B) -- node[right]{fusion} (D);
\draw[->] (C) -- node[below]{perturbation${}'$} (D);
\end{tikzpicture}
\caption{When is there a perturbation on the fusion product such that the diagram commutes?}
\label{fig:fusion}
\end{figure}  

However, if the situation for boundary flows can serve in any way as an indication to the defect situation, 
it may well be that there generically exists {\it some} flow that links the fusion product of some topological defect ${\cal I}_b$ with 
the original topologcial defect ${\cal I}_a$ in the UV with the fusion product of ${\cal I}_b$ with the original conformal defect in the IR. For boundaries, this was observed in~\cite{Graham:2003nc}. However, notice again that in particular the existence of defect changing operators will
greatly complicate a general analysis, as it does in the boundary case~\cite{Graham:2001pp}.

The situation where we perturb on the identity defect is the easiest to analyse.
The original perturbing field is actually a bulk field, which rather trivially also exists on all other elementary topological defects.
If we consider in particular the fusion of the identity with a group-like topological defect ${\cal I}_g$,\footnote{Group like topological defects  are associated to the elements $g$ of a symmetry group $G$, and satisfy ${\cal I}_g\star {\cal I}_{g'}={\cal I}_{gg'}$. They implement the symmetry of the theory.} all correlation functions of the perturbing operator will be unchanged. Figure~\ref{fig:grouplike} illutrates the reason for that statement. In this case we therefore indeed expect that fusion and perturbation commute.

\begin{figure}[th]
\centering
\begin{tikzpicture}
\node at (-1,0) {$\Big\langle$};
\node[draw,thick,circle,red,minimum size = 1.7cm] at (0,0) {};
\node[draw,cross=2.5,thick] at (.85,0) {};
\node[draw,cross=2.5,rotate=30,thick] at (-.425,.736) {};
\node[draw,cross=2.5,rotate=60,thick] at (-.425,-.736) {};
\node at (1,0) {$\Big\rangle$};
\node at (1.5,0){$=$};
\node at (2,0) {$\Big\langle$};
\node[draw,thick,circle,red,minimum size = 1cm] at (3,0) {};
\draw[red,dotted,thick] (3.5,0) -- (3.85,0);
\draw[red,dotted,thick] (3-.25,.433) -- (3-.425,.736);
\draw[red,dotted,thick] (3-.25,-.433) -- (3-.425,-.736);
\node[draw,cross=2.5,thick] at (3.85,0) {};
\node[draw,cross=2.5,rotate=30,thick] at (3-.425,.736) {};
\node[draw,cross=2.5,rotate=60,thick] at (3-.425,-.736) {};
\node at (4,0) {$\Big\rangle$};
\node at (4.5,0) {$=$};
\node at (5,0) {$\Big\langle$};
\node[draw,cross=2.5,thick] at (6.85,0) {};
\node[draw,cross=2.5,rotate=30,thick] at (6-.425,.736) {};
\node[draw,cross=2.5,rotate=60,thick] at (6-.425,-.736) {};
\node at (7,0) {$\Big\rangle$};
\draw[red, thick] (-.5,-1.5) -- (.5,-1.5);
\node at (.8,-1.55) {\textcolor{red}{$\mathcal{I}_g$}};
\draw[red, thick,dotted] (1.5,-1.5) -- (2.5,-1.5);
\node at (5,-1.53) {\textcolor{red}{$\mathcal{I}_g\mathcal{I}_g^\dagger = \mathcal{I}_g\mathcal{I}_{g^{-1}} = \mathcal{I}_{gg^{-1}} = id$}};
\end{tikzpicture}
\caption{Three-point function on group-like defects for bulk fields. One can shrink the defect circle to zero which in the vacuum expectation value gives a factor of $g_{I_g} = 1$. If the topological defect is not group like the dotted line is not simply the identity and can carry additional representations/fields that can appear in channels of higher point functions.  }
\label{fig:grouplike}
\end{figure}

A slightly more general situation occurs when the perturbation on the identity is compared to the perturbation by the same operator on a general elementary topological defect. 
In the previous section we have seen that the relevant perturbations on the identity generally trigger more relevant chiral perturbations (see \cite{Makabe:2017uch} for a concrete instance where the original perturbation would actually be irrelevant). 
This obviously changes the situation rather strongly. 
In the ${\cal N}=(1,1)$ Minimal Models, the perturbation by $\phi_{(1,3)(1,3)}$ triggers the much more relevant fields $\phi_{(1,3)(1,1)}$ and $\phi_{(1,1)(1,3)}$ on non-group-like elementary defects ${\cal I}_a$.
It would be very interesting to understand if there is any combination of these two operators (or others?) that yield a result which is the fusion product of ${\cal I}_a$ with the original almost-topological defect obtained from perturbing by $\phi_{(1,3)(1,3)}$ on the identity.


\section{Perturbative calculations}\label{sec:pertAnalysis}

In this section we calculate the change of the three properties of conformal defects introduced in section \ref{sec:pert}. We will consider perturbations of the identity defect. Most of the discussion treats the case of a spin-less deformation by a single field, which however can occasionally be adapted to the multi-parameter situation. 

The perturbative expansion is organized in powers of the small anomalous dimension $\delta = 1-\Delta $. 
From the proposals made in section \ref{sec:search}, the results hold for the $\phi_{(2,1)(2,1)}$ perturbation in the ${\cal N}=0$ Virasoro Minimal Models at low $k$, and in the $\mathcal{N}=(1,1)$  models for $\phi_{(2,1)(2,1)}$ at low $k$, and for $\phi_{(1,3)(1,3)}$ at high $k$. In the ${\cal N}=(2,2)$ models, only $\phi_{(2,0,0)(2,0,0)}$ at low $k$ is of that type. 
The last two examples have a non-vanishing self-OPE coefficient, such that we can use \eqref{eq:betaOPEscheme}. In the first examples we need to use \eqref{eq:betaCvanishes}.

\subsection{Entropy of the defect}

The entropy of the defect can be computed from the subleading contribution to the free energy on a very long torus \cite{Affleck:1991tk,Konechny:2014opa}. In general the free energy on a torus of length $L$ and circumference $2\pi$ with defect insertion takes the form 
\begin{equation}
F = \log Z = \frac{c}{12} L + A + \log g_\mathcal{I}^2 + \mathcal{O}(L)\,,
\end{equation}
where $A$ is a bulk spectrum dependent sub-leading contribution that also appears when there is no defect inserted and $g^2_\mathcal{I}$ is the square of the $g$-factor of the defect. It follows immediately that $g_\text{id} \equiv 1$. 

For a defect perturbation as discussed in section \ref{sec:pert}, the free energy in the large $L$ limit is  
\begin{equation}
 F-A = \frac{c}{12} L + \sum_{n=1}^{\infty} (\lambda)^n F_n + \dots\,,
\end{equation}
with 
\begin{align}\label{Fns}
 F_n &= \frac{1}{n!} \int_0^{2\pi} d\varphi_1 \dots d\varphi_n\, \langle \phi(i \varphi_1)\dots \phi(i \varphi_n)\rangle_{cyl.}\,,   
\end{align}
where $\phi$ is the perturbing field, $\lambda$ is its coupling, and the correlation function is that on a long cylinder.
The square of the $g$-factor of the perturbed defect is then given by $g_\mathcal{I}^2 = 1 +  \lambda^2 F_2 + \lambda^3 F_3 + \dots$. 
The contribution to first order in $\lambda$ vanishes because $\langle \phi(i\varphi)\rangle = 0$. 

By the folding trick, the $g$-theorem~\cite{PhysRevLett.67.161,Friedan:2003yc,Casini:2016fgb} 
for boundary RG flows will also hold for defects. 
The theorem states that the $g$-factor, which in a sense counts the elementary boundary degrees of freedom on the boundary or defect, decreases along relevant flows and does not change for exactly marginal deformations.

\subsubsection*{Second order in $\lambda$}

To second order in the renormalized coupling constant $\lambda$, the coefficient we have to compute is
\begin{align} \label{eq:EntInt}
 F_2 &= \frac{1}{2} \int_0^{2\pi} 
 d\varphi_1 d\varphi_2\, \frac{1}{(2\sin(\frac{\varphi_1-\varphi_2}{2}))^{2(1-\delta)}} \nonumber\\
 &= \int_0^{2\pi} d\varphi \, \frac{2\pi-\varphi}{(2\sin(\frac{\varphi}{2}))^{2(1-\delta)}}\,.
\end{align}
\noindent 
We regulate divergences by a cutoff angle $\epsilon$, and take the integral from $\epsilon$ to $2\pi-\epsilon$. In addition, we expand the integrand in $\delta$. The result is

\begin{equation}
 \left(\frac{2\pi }{\epsilon} \right) + \left(4\pi \frac{1+\log(\epsilon)}{\epsilon} - \pi^2\right) \delta + \left(4\pi \frac{2+2\log(\epsilon)+\log(\epsilon)^2}{\epsilon}-2\pi^2 \right) \delta^2 + \mathcal{O}(\delta^3)
\end{equation}
where we omit all terms that vanish for $\epsilon\rightarrow 0$. After regularization we thus get
\begin{equation}
 F_2 = -\pi^2\, \delta -2\pi^2\,\delta^2 + \mathcal{O}(\delta^3)\,.
\end{equation}

\noindent
In order to check this result, we can perform the integral analytically for $\delta>\frac{1}{2}$, where it produces the regular result 
\begin{equation}
 F_2 = \frac{4^\delta\pi^{3/2}\Gamma\left(\delta-\frac{1}{2}\right)}{2\Gamma(\delta)} \stackrel{\delta\to0}{\longrightarrow} -\pi^2\delta - 2\pi^2\delta^2 + O(\delta^3)\,.
\end{equation}

\noindent
Analytic continuation to small $\delta$  yields the same result as the cutoff-regularized integral. 

There is no contribution to $F_2$ for $\delta=0$, \textit{i.e.} if we perturb with a marginal field. This is what we expect from the $g$-theorem. 
For $\mathcal{C} = C_{\phi\phi}^\phi\neq0$, the subleading order in $\delta$ receives a correction from the third order in $\lambda$ at the IR fixed point.

\subsubsection*{Third order in $\lambda$}

The integral we have to solve is 

\begin{align}
 F_3 &= \frac{1}{6} \int_0^{2\pi} d\varphi_1 d\varphi_2d\varphi_3\, \frac{\mathcal{C}}{|z_{12}z_{23}z_{31}|^\Delta}\nonumber\\
     &= \frac{1}{6} \int_0^{2\pi} d\varphi_1 d\varphi_2d\varphi_3\, \frac{\mathcal{C}}{|z_{12}z_{23}z_{31}|} + \mathcal{O}(\delta)\,,
\end{align}
where $z_{ij} = z_i-z_j$, $z_i = e^{i\varphi_i}$ and we only keep terms of order $1$, \textit{s.t.} for $\mathcal{C}\neq 0$ we only keep results up to order $\delta^3$. The result of the integral now can be computed to be
\begin{align}
 \int_0^{2\pi} d^3\varphi\, \frac{1}{|z_{12}z_{23}z_{31}|} &= \int_0^{\pi} d^3\varphi\, \frac1{|\sin(\varphi_1-\varphi_2) \sin(\varphi_2-\varphi_3) \sin(\varphi_1-\varphi_3) |}\nonumber\\
							   &= 6 \int_{\varphi_1\ge\varphi_2\ge\varphi_3}d^3\varphi  \frac1{\sin(\varphi_1-\varphi_2) \sin(\varphi_2-\varphi_3) \sin(\varphi_1-\varphi_3) }\nonumber\\
							   &= 6 \int_0^\pi dx\int_0^{\pi-x} dy\, \frac{\pi-x-y}{\sin(x)\sin(x+y)\sin(y)}\nonumber\\
							   &= \frac{24 \pi \log(2)}{\epsilon} -3\pi^2\,,
\end{align}
where we again regulated divergences by $|\varphi_{ij}|\geq\epsilon$. 
After regularization, $F_3 = \frac{3\pi^2\mathcal{C}}{2} + \mathcal{O}(\delta)$. This result in particular shows that at third order in $\lambda$, only exactly marginal deformations, \textit{i.e.} perturbing operators with both $\delta$ and $\mathcal{C}$ equal to zero, do not change the $g$-factor.

With the result \eqref{eq:lC} for the value of the renormalized coupling in the IR, the $g$-factor for the perturbed defect becomes
\begin{equation}\label{gresult1}
g^2 = 1 -  \frac{5 \pi^2}{2\, \mathcal{C}^2} \,\delta^3+ \mathcal{O}(\delta^4)\,.
\end{equation}

\noindent
If we are in a situation where $C=0$, the value of the coupling constant in the IR is given by \eqref{eq:lC0} if ${\cal D}$ happens to be negative, in which case the first sub-leading contribution to $g$ is of order $\delta^2$, and the leading $\delta^0$ order of $F_4$ 
will contribute.

\subsection{Reflection at the defect}

Next we are going to compute the reflection coefficient as defined in \eqref{eq:ReflTrans}. The perturbative analysis on the identity defect is in many ways similar as in \cite{Brunner:2015vva}. A perturbative analysis of the reflectivity for the two parameter deformations introduced in \cite{Kormos:2009sk} was done in \cite{Makabe:2017uch}.

The initial topological defect is again wrapped around a cylinder of circumference $2\pi$. We define the unitary matrix
\begin{equation}
 R = \begin{pmatrix}
      \langle \mathcal{I}\, T\bar{T}\rangle & \langle T\, \mathcal{I}\, T \rangle\\
      \langle\tilde T\,\mathcal{I} \,\tilde T \rangle & \langle T\bar{T}\,\mathcal{I}\rangle
     \end{pmatrix} \equiv
     \begin{pmatrix}
      R_{11} & R_{12}\\
      R_{21} & R_{22}
     \end{pmatrix}\,,
\end{equation}
where the insertion points of the energy momentum tensor on the plane are at some value $z$ and at the reflected point $w = 1/z^*$.
The reflection coefficient is then given by 
\begin{equation}
\mathcal{R} = \frac{R_{11}+R_{22}}{\mathcal{N}}\,,
\end{equation}
where $\mathcal{N} = \sum_{ij} R_{ij}$ is a normalization. The situation is invariant under reflection at the defect, such that $R_{11}= R_{22}$, and $\mathcal{R}$ is independent of the actual insertion $z$, and we push  $z,\,w$ to infinity (or zero, respectively). In this limit we have $\mathcal{N} = g\, c$, where $g$ is the defect entropy.  The quantity we want to compute in perturbation theory is 
\begin{equation}
 R_{22} = \lambda^2 R_{22}^{(2)}+\lambda^3 R_{22}^{(3)} \lambda^4 R_{22}^{(4)} + \mathcal{O}(\lambda^5)
\end{equation}
for a spinless perturbing operator $\phi$, where 
\begin{equation}
R^{(n)} = \frac{1}{n!} \int_0^{2\pi} d^n\varphi \,\langle T(\infty)\bar{T}(\infty) \mathcal{I}\phi(e^{i\varphi_1})\dots \phi(e^{i\varphi_1}) \rangle\,.
\end{equation}

\noindent
It is clear that there is no zeroth or first order contribution, because the two correlators $\langle T(z)\bar{T}(\bar{z})\rangle$ and $\langle T(z)\bar{T}(\bar{z})\phi(e^{i\varphi})\rangle$ vanish. From now on we set $\mathcal{I} = \text{id}$ in the calculation, because any other $\mathcal{I}$ just contributes an overall factor $g_\mathcal{I}$ that is removed by the above normalization. 

\subsubsection*{Second order in $\lambda$}

Using conformal Ward identities one can show that $\langle T(\infty)\bar{T}(\infty) \phi(z_1) \phi(z_2)\rangle = \frac{\Delta^2}{4}|z_{21}|^{2+2\delta}$. We therefore obtain
\begin{align}
 R_{11}^{(2)} &= \frac{\Delta^2}{8} \int_0^{2\pi} d\varphi_1 d\varphi_2 \,|e^{i\varphi_1}-e^{i\varphi_2}|^{2+2\delta}  = \frac{\Delta^2}{8} \frac{4^{2+\delta} \pi^{3/2} \Gamma\left(\frac{3}{2}+ \delta\right)}{\Gamma(2+\delta)}\nonumber\\
 & = \Delta^2 \left(\pi^2 + \pi^2 \delta +\mathcal{O}(\delta^2)\right) = \pi^2 - \pi^2\delta + \mathcal{O}(\delta^2)\,. \label{eq:R2}
\end{align}

\noindent
We observe that even marginal deformations lead to a change in the reflection coefficient. That marginal deformations lead to an increased reflection has been observed in the Ising model with its three classes of conformal defects that derive from exactly marginal deformations of the three topological defects~\cite{Oshikawa:1996dj}. 

\subsubsection*{Third order in $\lambda$}

Again using the conformal Ward identities one can show that 
$$\langle T(\infty)\bar{T}(\infty) \phi(e^{i\varphi_1}) \phi(e^{i\varphi_2}) \phi(e^{i\varphi_3})\rangle = \frac{\mathcal{C}\Delta^2}{4}\frac{|z_1^2+z_2^2+z_3^2-z_1z_2-z_2z_3-z_3z_1|^2}{|z_{12}z_{23}z_{31}|^\Delta}\,.$$

\noindent 
Using this result we find
\begin{align}
 R_{11}^{(3)} &= \frac{\mathcal{C}\Delta^2}{24} \int_0^{2\pi} d\varphi_1d\varphi_2d\varphi_3 \left|\frac{1}{z_{12}}+\frac{1}{z_{23}}+\frac{1}{z_{31}}\right|^2 |z_{12}z_{23}z_{31}|^{1+\delta}\nonumber\\
	      &= \frac{\mathcal{C}}{24} \int_0^{2\pi} d\varphi_1d\varphi_2d\varphi_3 \left|\frac{1}{z_{12}}+\frac{1}{z_{23}}+\frac{1}{z_{31}}\right|^2 |z_{12}z_{23}z_{31}|+\mathcal{O}(\delta)\nonumber\\
	      &= \frac{\mathcal{C}}{24} \int_0^{2\pi} d\varphi_1d\varphi_2d\varphi_3\,\left( 3\frac{|z_{23}z_{31}|}{|z_{12}|} + 6 \sqrt{\frac{\bar z_{12} z_{23}}{ z_{12} \bar z_{23}}} |z_{31}| \right)\nonumber\\
	      &=  \frac{\mathcal{C}}{24} \left( \int_0^{2\pi} d\varphi_1d\varphi_2d\varphi_3\,3\frac{|z_{23}z_{31}|}{|z_{12}|} + 48 \pi^2 \right)\,.
\end{align}
We only keep the leading order in $\delta$. The remaining integral can be massaged further to yield
\begin{align}
 \int_0^{2\pi} d\varphi_1d\varphi_2d\varphi_3\,3\frac{|z_{23}z_{31}|}{|z_{12}|} &= 48 \int_0^{\pi} d\tilde\varphi_1d\tilde\varphi_2d\tilde\varphi_3 \left|\frac{\sin\left(\tilde\varphi_1-\tilde\varphi_3\right)\sin\left(\tilde\varphi_2-\tilde\varphi_3\right)}{\sin\left(\tilde\varphi_1-\tilde\varphi_2\right)}\right|\nonumber\\
 &=288 \int_{\varphi_1\ge\varphi_2\ge\varphi_3} d\tilde\varphi_1d\tilde\varphi_2d\tilde\varphi_3 \frac{\sin\left(\tilde\varphi_1-\tilde\varphi_3\right)\sin\left(\tilde\varphi_2-\tilde\varphi_3\right)}{\sin\left(\tilde\varphi_1-\tilde\varphi_2\right)}\nonumber\\
 &= 288 \int_0^\pi dx \int_0^{\pi-x} dy \, (\pi-x-y) \frac{\sin(y)\sin(x+y)}{\sin(x)}\nonumber\\
 &= 72 \int_0^\pi dx \,(\pi-x)(1+(\pi-x)\cot(x))\nonumber\\
 &= 12 (3\pi^2-6\pi^2\log(2)-6\pi^2\log(\tilde\epsilon)) + \mathcal{O}(\tilde\epsilon) \nonumber\\
 &=12 (3\pi^2-6\pi^2\log(\epsilon)) + \mathcal{O}(\epsilon) \,,
\end{align}
where $\epsilon$ is a angular space UV cutoff.  Assuming that suitable counterterms will remove the divergent terms, the result is 
\begin{equation}
 R^{(3)}_{11} = \frac{7\pi^2 \mathcal{C}}{2}\,.
\end{equation}

\noindent
For $\mathcal{C}\neq0$ we can use \eqref{eq:lC}, and the reflection coefficient becomes
\begin{equation}\label{Rresult1}
 \mathcal{R} = \frac{2 \pi^2}{c\,\mathcal{C}^2} \, \delta^2 +\frac{\left(5\mathcal{C}^2 -4{\cal D}\right)\pi^2}{ c\,\mathcal{C}^4} \,\delta^3 + \mathcal{O}\left(\delta^4\right)\,.
\end{equation}

\noindent 
In the case where $\mathcal{C}=0$, \textit{i.e.} when we have to use \eqref{eq:lC0}, our calculations still allow us to give the result up to the first non-vanishing order:\footnote{Contrary to the situation for the $g$-factor, we do not need the result from fourth order in $\lambda$ here, because $\mathcal{R}^{(2)}_{1,1}$ contributes also to order $\delta^0$.}
\begin{equation}\label{Rresult2}
\mathcal{R} = \frac{2\pi^2}{c\,\mathcal{D}} \delta + \mathcal{O}(\delta^{3/2})\,.
\end{equation}

\subsection{Entanglement entropy through the defect}

As briefly discussed in section \ref{sec:properties}, one way to compute the entanglement entropy through a defect is to first compute the free energy of the theory on a torus with $2K$ defect insertions. Similarly to the case of the $g$ factor, the free energy 
can be written as
\begin{equation}
 \log Z_K = \frac{c \pi^2}{6 d K} + \lambda^2 F_{2,K} + \mathcal{O}(\lambda^3)\,,
\end{equation}
where $d \equiv \frac{2\pi^2}{\log\frac{L}{\epsilon}}$ and 
\begin{equation}\label{eq:EEF2}
 F_{2,K} = \frac{1}{2} \sum_{n_1,n_2=1}^{2K} \int_{\log\epsilon}^{\log L}dx_1dx_2\, \langle \phi(x_1 + in_1\pi ) \phi(x_2 +in_2\pi) \rangle_\text{cyl.,K}\,. 
\end{equation}

\noindent 
In the last expression, the integration variables run along the $2K$ parallel defect lines, running at equal distances along the cylinder.  
In the following we will only need to keep contributions from  $F_{2,K}$ to leading 
order in $\delta$.

We perform the integral on the plane with coordinate $w = e^{z/K}\equiv R e^{i \phi}$. For $n=n_1-n_2$ not a multiple of $K$
we obtain
\begin{equation}
I(n)\equiv\int_{\epsilon^{1/K}}^{L^{1/K}}  \frac{dR_1dR_2}{R_1^2+R_2^2-2 R_1R_2 \cos\left(\frac{n\pi}{K}\right)} = \frac{\pi}{K \sin\left(\frac{\pi n}{K}\right)} \left(1-\frac{n}{K}\right) \log\frac{L}{\epsilon}\,.
\end{equation}

\noindent
which is also the result for $-n, 2K-n,$ and $n-2K$. The result for $n=0$ and $n=K$ can be obtained by taking the respective limit
\begin{align}
I(n\to 0) &= \frac{\log\frac{L}{\epsilon}}{n} -\frac{\log\frac{L}{\epsilon}}{k}\,,\\
I(n\to K) &= \frac{\log\frac{L}{\epsilon}}{k}\,.
\end{align}

\noindent
At $n\to0$ we can see the expected divergence that has to be regularized. Since any $n = n_1-n_2$ appears $K$ times in the above sum, we find
\begin{equation}
 F_{2,K} =  2 \pi \log\frac{L}{\epsilon}\sum_{n=1}^{K-1}\csc\left(\frac{n\pi}{K}\right)\left(1-\frac{n}{K}\right) \,.
\end{equation}

\noindent
For the entanglement entropy we have
\begin{equation}
 E = \lim_{K\to1} (1-\partial_K) \log Z_K = E^{(0)} + \lambda^2 E^{(2)} + \mathcal{O}(\lambda^2)\,,
\end{equation}
with
\begin{equation}
 E^{(2)} =  F_{2,1} - F'_{2,1}\,,
\end{equation}
where a prime denotes a derivative with respect to $K$. We evaluate $E^{(2)} $ by computing it numerically for values $K = 1,\dots,K^\text{max}$, and then use an interpolating function to compute $F'_{2,1}$. The numerics suggest the result   
\begin{equation}
E^{(2)} = -\frac{\pi^2}{4}\log\frac{L}{\epsilon}\,,
\end{equation}
such that the entanglement entropy through the perturbed defect up to order $\lambda^2$ is given by\footnote{The fact that we obtain $c/6$ instead of the familiar $c/3$ in leading order is due to the special choice of reducing on a half-line \cite{sakai_entanglement_2008,Brehm:2015lja,Brehm:2015plf}. Intuitively it can be understood from the area law: Our entangling surface consists of only the origin, rather than of two points.}
\begin{equation}
E = \frac{c}{6}\left(1  - \frac{3\pi^2}{2c}\lambda^2 \right)\log\frac{L}{\epsilon}\,.
\end{equation}
\noindent
Let us assume that the entanglement entropy through the defect directly depends on reflection at the defect. Then, by using \eqref{eq:R2}, the above result up to order $\lambda^2$ can be written as 
\begin{equation}\label{E-Rrelation}
  E = \frac{c}{6}\left(1  - \frac{3}{4} \mathcal{R} \right)\log\frac{L}{\epsilon}\,.
\end{equation}

\noindent 
This matches the result for defects in the critical Ising model~\cite{Brehm:2015lja}, but also with the 
result of~\cite{sakai_entanglement_2008}\footnote{In this case, the match holds through the relation $s^2=1-{\cal R}$ \cite{quella_reflection_2007}}.
In both cases, which we therefore conjecture to hold universally for the perturbations considered here. For the Ising model, the entanglement entropy through defects with arbitrary reflectivity $\mathcal{R}\in [0,1]$ is given by \cite{Brehm:2015lja} 
\begin{equation}
E_\text{Ising}(\mathcal{R}) = \frac{c}{6}\, \sigma(s) \log\frac{L}{\epsilon}\,
\end{equation}
with $s = \sqrt{1-\mathcal{R}}$ and 
\begin{align}
\sigma(s) &= s-1 - \frac{6}{\pi^2}\left((s+1)\log(s+1)\log s +(s-1)\text{Li}_2(1-s)+(s+1)\text{Li}_2(-s)\right)\nonumber\\
&= 1 - \frac{3}{4} \mathcal{R} + \mathcal{O}(\mathcal{R}^2)\,.
\end{align}
\noindent
Our result again confirms the expectation that the entanglement entropy through a topological defect is reduced by almost marginal deformations. Using the value \eqref{eq:lC} for the coupling in the IR, the perturbative change in entanglement entropy is
\begin{equation}\label{Eresult}
 \Delta E = - \frac{3 \pi^2 \delta^2}{2 c \mathcal{C}^2} \log\frac{L}{\epsilon} + \mathcal{O}(\delta^2)\,.
\end{equation}
In the case where $\mathcal{C}=0$, the value \eqref{eq:lC0} of the coupling in the IR leads to the result $\Delta E = \frac{3 \pi^2 \delta }{2 c\cdot \mathcal{D}} \log\frac{L}{\epsilon}+ \mathcal{O}(\delta^{3/2})$.

\section{Conclusion}

Our first goal in this paper was to present candidate perturbations of elementary topological defects,
tractable by perturbative methods, within the well-understood coset 
models~\mbox{\eqref{eq:CosetN0} -- \eqref{eq:CosetN2}}. We have found the operators 
displayed in Table~\ref{tab:fields} to yield candidates for such RG flows. 
They include, in particular, the $\mathcal{N}=1$ supersymmetric version of the perturbation considered in~\cite{Kormos:2009sk}.
 
Our second goal was to take some steps towards the understanding of the IR limit of these perturbations, by computing the entropy, reflectivity and entanglement entropy through the defect that results from perturbing the identity by a single spinless, almost marginal field. 
This leads to the results~\eqref{gresult1}, \eqref{Rresult1}, \eqref{Rresult2}, and~\eqref{Eresult}. 
Furthermore, the relation \eqref{E-Rrelation} confirms the idea that the entanglement entropy through the defect is at least perturbatively tied to the transmission of energy and momentum.
 
Obviously we have merely scratched the surface of the study of near-topological defects. 
Our perturbation theory is for example only valid for very particular fields and only for perturbations on the identity defect. However, generalization should mostly be straight forward. Our perturbative results can be seen as special case that give an intuition on how defects physically behave. 
It is unfortunate that in particular the minimal models we investigates do not have many examples for which we can directly apply our results. However, they are in fact universal and can be applied to any unitary CFT that contain at least one field that behaves as we demanded in section \ref{sec:defectPurt}. 

We point out that there is a much more complete picture for the case of boundary RG flows, especially in the framework of coset models (see {\it e.g.}~\cite{Recknagel:2000ri,Graham:2001pp,Fredenhagen:2003xf}). 
In particular, \cite{Fredenhagen:2003xf} gives a rather conclusive boundary flow analysis by generalising the ``absorption of boundary spin'' method of \cite{Affleck:1990by}. Since our flows can be cast as boundary flows of coset models after the folding trick, it is quite possible that these methods also yield information on the perturbative flows we consider here, although we have not attempted to work this out. 
 
\acknowledgments

We really like to thank Cornelius Schmidt-Colinet who helped a lot in the development of the project. We also like to thank Stefan Theisen, Diptarka Das, Gerard Watts, and Ingo Runkel for helpful discussions and/or suggestions for improvement of the manuscript. 





\bibliographystyle{JHEP}
\bibliography{lit}
\end{document}